\documentstyle{article}

\begin{document}

 \noindent {\Large {\bf Direct numerical solution  of
     the two-particle Lippmann-Schwinger equation in coordinate space
     using the  multi-variable Nystrom method}}

 \vspace*{0.5cm}

 \noindent Zeki C. Kuruo\u{g}lu\\
  Department of Chemistry, \\
  Bilkent University,\\
  06800 Bilkent, Ankara,Turkey\\
  E-mail: kuruoglu@bilkent.edu.tr\\

\vspace*{1.5cm}

 \noindent {\bf Abstract}
   Direct numerical solution of the coordinate-space integral-equation version of
the two-particle Lippmann Schwinger (LS) equation  is considered
as a means of avoiding the shortcomings of partial-wave expansion
at high energies and in the context of few-body problems.
Upon the regularization of the singular kernel of the
three-dimensional LS equation by a subtraction technique,
a three-variate quadrature rule is used to solve the resulting nonsingular integral
equation. To avoid the computational burden of discretizing three variables,
advantage is taken of the fact that, for central potentials, azimuthal angle can be
integrated out leaving a two-variable reduced integral equation.
Although the singularity in the the kernel
of the two-variable integral equation is weaker than that of the three-dimensional equation,
it nevertheless requires careful handling for quadrature discretization to be applicable.
A  regularization method for the kernel of the two-variable integral equation
is derived from the treatment of the singularity in the three-dimensional equation.
A quadrature rule constructed as the  direct-product of  single-variable quadrature rules
 for radial distance and polar angle
is used to discretize the two-variable integral equation.
 These two- and three-variable methods are tested on a model nucleon-nucleon potential.
The results show that Nystrom method for  the coordinate-space LS equation
 compares favorably  in terms of its ease of implementation and effectiveness
 with the Nystrom method for the momentum-space  version of the LS equation .

\vspace*{1cm}

\noindent {\bf 1  Introduction } \\

\noindent Expansion in angular-momentum states has hitherto been the usual ansatz
for computational approaches to   quantum-mechanical scattering problems.
 However, a critical re-assesment  of this strategy
 has occured during  recent years,
 especially for high-energy collisions [1] and
 within  the context of few-body problems [2].
It has been realized that even for two-body problems involving central potentials,
where the advantage due to the decoupling of partial wave equations is manifest,
the partial wave expansion might loose its practical edge in high energies and for use
in few-body calculations employing Faddeev-Yakubovski-type equations.
Although  the
  scattering amplitudes for most potentials are rather smooth,
   partial wave amplitudes may show oscillatory behavior. Similarly, the off-shell two-body $T$-matrix
   has usually simple structure whereas partial wave components
   might strongly oscillate.
   Under such circumstances,  the partial wave expansion involving
   an excessively large number of
   partial waves may be computationally impractical or even unreliable .

      These observations suggest that, to treat two-particle scattering at high
      energies and  within the context of few-particle dynamics,
      direct multi-variable methods without recourse to
      expansions over angular momentum states might be more appropriate.
        Towards this end, multivariable methods has been
			investigated for the solution of the
			 multi-variable LS equation in the momentum space [ 3-13].
			For example in Refs. [11, 12] , we have considered
			multivariable implementations of Schwinger variational
            and Bateman methods for two-body LS equation in momentum space.
			Significant progress has also been reported
             on the formal and computational aspects of solving
            the three-particle momentum-space Faddeev equations
			directly as 5-variable problems
			without invoking angular momentum decomposition [2, 14, 15].

  Calculational  schemes based on momentum-space LS equations
  dominate the literature for two-body scattering computations,
	as exemplified in [3, 5-13].
  Coordinate-space version of the LS equation   have received relatively
  less attention as a computational vehicle, although the coordinate-space partial-wave
   LS equation    has been employed in connection with
   various types of Schwinger variational methods [16-17].
	As far as the present author
   is able to ascertain, the direct numerical solution  of the three-dimensional  coordinate-space LS
   equation for the transition operator does not appear to have been  reported  before.
	 Presumably this is due to the
	 singularity of the free Green's function $G_0\, ( {\bf r} ,{\bf r'}) $ in the kernel of the LS equation.
   The most straightforward approach to solve an integral equation    is
	 the so-called Nystrom method[18] in which  the integral equation is converted to
    a system of linear equations by approximating
    the  integral by a quadrature. However, as the Green's function
  $G_0\, ( {\bf r} ,{\bf r'}) $ in the kernel of the LS equation  becomes
  singular as $  {\bf r}\,  \rightarrow \,  {\bf r'}$,
  the Nystrom method  can be applied only after the kernel of the LS equation
	is regularized. Similar singularities also occur in integral-equation formulations
	of electromagnetic scattering.
	In this article,  a subtraction technique commonly
	 used in computational electromagnetics [19,20]  is adopted to the three-dimensional LS equation.
	This subtraction scheme regularizes the singular kernel of the three-dimensional LS equation 	
	 and  brings it into a form apropriate for the application of the Nystrom method.
	
    Employing a direct-product quadrature rule
		with  a quadrature mesh $\{  {\bf r}_{\alpha}\, ,   \,
     \alpha =1,2,...,N_Q \}$
     for integration over the computational domain
     in coordinate space, Nystrom method yields  a system of $N_Q$ linear equations for
     the (mixed representation) matrix elements $ < {\bf r}_{\alpha} |T(E+)|{\bf q_0}>\, $,
    where $|{\bf q_0}> $ is the initial state
       with relative momentum ${\bf q_0}$ and energy $E\, =\, |{\bf q}_0|^2/2\mu \, $.
       The  momentum-space representation $<{\bf q}|T(E+)|{\bf q_0}>$ is then obtained
       from  $<{\bf r}|T(E+)|{\bf q_0}>$
       by the same  three-dimensional quadrature used in the Nystrom method.

    In our implementation of the
			three-dimensional Nystrom method for model potentials,
			3-4 digit accuracy for scattering amplitude
			could be obtained with
			$N_Q$ in the order of $30-40$ thousand quadrature points
			(involving 60-100 points in $r$, and $16-20$
			points in $\theta$ and $\phi$  angles each.
			As usual with   direct-product approaches to multi-variable problems,
			the curse of dimensionality makes
			 reaching higher level of accuracy a formidable task. Fortunately,
			however, for central potentials,
			number of variables can be reduced by one.
			Using the fact that  $G_0\, ( {\bf r} ,{\bf r'}) $,  and $<{\bf r}|T(E+)|{\bf q+0}>$
			for central potentials,
   depend on azimuthal angles only via the cosine  of the difference
   of the azimuthal angles of the vectors involved, the azimuthal-angle dependence
	 in the LS equation can be eliminated. By integration over the azimuthal angle,
	 the three-dimensional equation reduces to
   a two-variable LS equation  for the reduced matrix element
$<r\theta|T(E)|q_0\theta_{q_0}>$, where $r=|{\bf r}|$ , $q_0=|{\bf q_0}|$ and
$\theta $ and $\theta_{q_0}$ are the polar angels associated
with vectors ${\bf r}$ and ${\bf q_0}$, respectively.  This reduced integral equation
 can be considered as the integral-equation counterpart of the
 two-variable differential equation that was solved  in Ref. [4]
 using the finite-element method to avoid partial wave expansion.

Integration over the azimuthal angle
weakens the singularity of the original Green's function, but the new reduced kernel
still  requires careful handling.  We show that the  regularized non-singular
three-dimensional integral  equation  can be reduced
to obtain a regularized two-variable equation
which is in a form ready for the quadrature discretization.  Constructing
 a direct-product quadrature scheme by using $N_r$ points in $r$ and
	 $N_{\theta}$ points in $\theta$,  Nystrom solution of the reduced LS equation yield a linear system of
	$N_rN_{\theta}$ equations which can be solved routinely
	in commonly available computational platforms.

             Plan of this article is as follows: In Sect. 2, we discuss the reduction of  the
              three-dimensional  LS equation
            into a two-variable integral equation.
						Sect. 3 discusses the subtraction scheme for the removal of the singularity
						from the kernels of the three dimensional and  reduced forms of the LS equation.
               In Sect. 4,   the details of the computational implementation and results of
							 calculations for the model potential are presented.
                In Sect. 5 we summarize our conclusions.\\

 \noindent {\bf 2 Lippmann-Schwinger Equation } \\

\noindent We consider the two-particle scattering problem for a central interparticle potential.
 Working in  the center-of-mass frame, the relative momentum states are denoted by $|\bf q>$ and
	the relative position states by $|\bf r>$, with the normalizations
	$ <{\bf r}|{\bf r'}>\, = \, \delta ({\bf r} -{\bf r'}>$,  $\  <{\bf q}|{\bf q'}>\, =\,  \delta ({\bf q} -{\bf q'}>$,
	and $<{\bf r}|{\bf q}> \, =\,  e^{i {\bf r}\cdot {\bf q} }/(2\pi)^{3/2}\, $.
	We will set $\hbar =1$ throughout this article.

Basic equation for the description of two-particle scattering is
the Lippman-Schwinger equation for the two-particle transition
operator $T(z)$:
\begin{equation}
T(z)\ =\ V\ + V\, G_0(z)\, T(z)\ ,
\end{equation}
 where $V$ is the interaction potential between two particles,
 $G_0\,=(z-H_0)^{-1}$, with  $z$  being the (complex) energy of the two-particle system.
  For on-shell scattering,  $z=E+i0$ with $E=q_0^2/2\mu$, where $\mu$ is the reduced mass.
  Using a mixed representation, the matrix elements $T({\bf r},{\bf q}_0 )(\equiv
  <{\bf r}|T(E+i0)|{\bf q}_0>$)
  satisfy  the three-dimensional integral equation
 \begin{equation}
 T ({\bf r}, {\bf q}_0  )\ =
 \ V (r)\, <{\bf r}\, | {\bf q}_0 >\, + \ V(r) \,  \int \, {\mbox{d} }{\bf r'} \,
 \, G_0({\bf r}, {\bf r'}) \ T ({\bf r'}, {\bf q}_0 )\, ,
\end{equation}
where $G_0({\bf r}, {\bf r'})$ is the free Green's function, viz.,

\begin{equation}
 G_0({\bf r}, {\bf r'}) \,  =\, <{\bf r}|G_0(E+i0)|{\bf r'}> \,  = \,
 -\frac{\mu}{2\pi} \, \frac{e^{iq_0|{\bf r}-{\bf r'}|} } { |{\bf r}-{\bf r'}|}\, .
\end{equation}
Note that the $T ({\bf r}, {\bf q}_0 )$ is closely related to the scattering wave function
 $\psi_{\bf q_0}({\bf r})$, viz.,
 \begin{equation}
 T ({\bf r}, {\bf q}_0 )\, =\,  V(r) \psi_{{\bf q_0}}({\bf r})\, ,
\end{equation}
where $\psi_{{\bf q_0}}$ is the solution of the LS equation for the wave function
\begin{equation}
 \psi_{{\bf q_0}}({\bf r})\,  =
 \, <{\bf r}\, | {\bf q}_0 >\, + \,  \int \, {\mbox{d} }{\bf r'} \,
 \, G_0({\bf r}, {\bf r'}) \, V(r')\, \psi_{{\bf q_0}}({\bf r'}) \, .
\end{equation}

Since $V(r)$ vanishes as $r$ gets sufficiently large,
solving Eq.2 for the amplitude $T ({\bf r}, {\bf q}_0 )$
proves to be much more convenient computationaly than directly
working with Eq. 5 for the wave function.
The momentum-space matrix elements
$T\,({\bf q}, \, {\bf q_0} )\, (\equiv
 <{\bf q}|T(E+i0)|{\bf q_0}>\, )$
of the transition operator
 can be calculated from the solution of Eq.2 via  a quadrature:
 \begin {equation}
 T({\bf q}, \, {\bf q_0} )\,= \
 \,  \int \, d{\bf r}\, \, <{\bf q}|{\bf r} >\,
 T ({\bf r},\, {\bf q}_0 ).
 \end{equation}

 We will denote
 the polar and azimuthal angles of the position  vector ${\bf r}\, $
 by $\theta  \, $
 and $\phi\, $, and those of the momentum vector $\bf q$
 by $\theta_q  \, $
 and $\phi_q\, $, respectively. We will also use the notation $x$ for $\cos \theta$,
$s$  for $\sin \theta$,
  $x_q$ for  $\cos \, \theta_q$, and $s_q$ for $\sin \theta_q$.
 Since $|{\bf r}-{\bf r'}|= {\sqrt {r^2 + r'^2 -2rr'x_{rr'}} }$ with
 $x_{rr'}=xx'+ ss'\, cos\, ( \phi-\phi')$,  dependence of
  $G_0({\bf r}, \, {\bf r'} )$
   on azimuthal angles is only through the
  difference $\phi - \phi'$. Similarly
  $T({\bf r}, \, {\bf q_0})$ for central potentials depend on $r$, $q_0$, and
  $x_{rq_0}$\, . Here $x_{rq_0}$ is the cosine of the angle
	between vectors $\bf r$ and ${\bf q_0}$, i.e.,
	$x_{rq_0} \, =\,  {\hat {\bf r} }\cdot {\hat {\bf q_0}} $.

  Towards  the elimination of the azimuthal angles,
  we introduce the   states $\, |r x> $ and  $\, |q x_q>\, $  via
  \begin {eqnarray}
  |rx>\  =\
  (2\pi)^{-1/2}\, \int_{0}^{2\pi}\, d\phi \,\, |{\bf r}>\, =
   \, (2\pi)^{-1/2}\, \int_{0}^{2\pi}\, d\phi \,\, |r\theta\phi>\,  , \\
  |q x_q>\   =\
   (2\pi)^{-1/2}\, \int_{0}^{2\pi}\, d\phi_q \,\, |{\bf q}>\, =
   \, (2\pi)^{-1/2}\, \int_{0}^{2\pi}\, d\phi_q \,\, |q\theta_q\phi_q>\, .
  \end{eqnarray}
 We next introduce reduced matrix
  elements of $G_0$ and $T$ via
\begin{eqnarray}
 {\hat  G_0}(r,x;r',x') \ = \   <rx|G_0|r'x'> \  = \
  (2\pi)^{-1}\int_0^{2\pi} \, d\phi\,\, \int_0^{2\pi} \, d\phi'\,\,  G_0({\bf r},{\bf r}'),\\
 {\hat  T}(r,x;q_0,x_{q_0}) \ = \   <rx|T|q_0x_{q_0}> \  = \
  (2\pi)^{-1}\int_0^{2\pi} \, d\phi\,\, \int_0^{2\pi} \, d\phi_{q_0}\,\,  T({\bf r},{\bf q_0}).
 \end{eqnarray}
 Note that operators, matrix elements and other quantities associated with  the two-variable
 representation will be distinguished from those of three-dimensional representation
 by a caret over the symbol.

Since $G_0({\bf r},{\bf r'})$ and $T({\bf r},{\bf q_0})$ depend on azimuthal angles
only through the
differences $\phi - \phi' $ and $\phi - \phi_{q_0}$, respectively,  integration over one of the
azimuthal angles  can be carried out to obtain
\begin{eqnarray}
  {\hat  G_0}(r,x;r',x')\ =\ \int_0^{2\pi} \, d\phi\,\, G_0({\bf r},{\bf r'}) \ =\
   \int_0^{2\pi} \, d\phi'\,\, G_0({\bf r},{\bf r}')\, , \\
 {\hat  T}(r,x;q_0,x_{q_0})\ =\ \int_0^{2\pi} \, d\phi\,\, T({\bf r},{\bf q_0}) \ =\
   \int_0^{2\pi} \, d\phi_{q_0}\,\, T({\bf r},{\bf q_0})\,  .
  \end{eqnarray}
\noindent  Note that , in Eq. 11, the first integral on the right-hand side  is
independent of the azimuthal angle $\phi'$ of ${\bf r'}.$,
while the second integral is independent of  $\phi$.
Integrating both sides of Eq. (2) over $\phi$ and $\phi_{q_0}$,
interchanging  the order of integration over $\phi$ and
$\phi'$ and then  making use of  Eqs. 11 and 12, we obtain
 the reduced two-variable LS equation
\begin{eqnarray}
 {\hat T}(r,x;q_0,x_0)&  =  & V(r)<rx|q_0x_{q_0}>    \  \nonumber \\
        \,& \ & +\,    V(r) \int_{0}^{\infty} r'^2 dr' \int_{-1}^{1}  dx' \,
   {\hat G}_0(r,x;r',x')\,  {\hat T}(r',x';q_0,x_{q_0}) \, ,   \nonumber \\
	    \ & \  & \
  \end{eqnarray}	
	where
	\begin{equation}
	<rx|q_0x_{q_0}> \,\,   =\, \,  \int_0^{2\pi} \, d\phi \, <{\bf r}| {\bf q_0> }\, =
	 \,  (2\pi)^{-1/2}\, e^{iq_0rxx_{q_0} }\, J_0\, (q_0rss_{q_0}) \,  ,
	\end{equation}
	with $J_0$ denoting the zeroth order Bessel function. We write Eq. 7 in operator form as
  \begin{equation}
\hat T\ = \hat V + \hat V \hat G_0 \hat T \, ,
\end{equation}
which is to be understood as an operator equation in the  space
of  two-variable functions (of $r$ and $x$).
On the other hand, the reduced version of Eq. 5 reads
\begin{equation}
{\hat T}(q,x_q;q_0,x_{q_0}) \, =\, \int \, r^2 \, dr \, \int_{-1}^{1} \, dx \,
 <qx_q|rx> {\hat T}( r,x;q_0,x_{q_0} ).
\end{equation}

 As discussed in [11], for an initial momentum vector ${\bf q}_0$
  along the z axis and a general final momentum vector $\bf q$,  the transition matrix element
	${\bf q}|T|q_0\hat {\bf z}>$ is given by
  \begin {displaymath}
  <{\bf q}|T|q_0\hat {\bf z}>\,  =\,  (2\pi)^{-1}\,  T(q,x_q\, ;q_0, 1)\, . \nonumber
  \end{displaymath} \\

  \noindent {\bf 2 Regularization of the singular kernels and the Nystrom method} \\

 \noindent The standard method for the numerical solution of non-singular integral equations
 is the Nystrom method, in which the integrals  are replaced by sums
via suitable quadrature rules and  the resulting equations are collocated at the qudrature points
 However, singular nature of $G_0({\bf r}, {\bf r'}) $ does not allow
the direct application of Nystrom method to Eq. (2)
in full three-dimensional approach or Eq. (7) the in two-variable version . We first recast
the singular Green's function as a sum of non-singular and
singular parts by subtracting and adding an (analytically) integrable  singular  term:
\begin {eqnarray}
 G_0({\bf r}, {\bf r'}) \,  = \,
 -\frac{\mu}{2\pi} \,  {\Big \{ } \,  \frac{e^{iq_0|{\bf r}-{\bf r'}|} } { |{\bf r}-{\bf r'}|}\, - \, \frac{1 } { |{\bf r}-{\bf r'}|}\,
{\Big \} }  \
   \, -\frac{\mu}{2\pi} \,   \, \frac{1 } { |{\bf r}-{\bf r'}|}\,
\end {eqnarray}
\noindent where the term within curly brackets is
 no longer singular as ${\bf r'}\rightarrow {\bf r} $, while the last term
 is analytically integrable over ${\bf r'}$ for fixed ${\bf r}$.

Using this  splitting in Eq. (2 ), we obtain
\begin{eqnarray}
   T ({\bf r}, {\bf q}_0  )\, & =  & \,
   V (r)\, <{\bf r}\, | {\bf q}_0 >\,
  - \, \frac{\mu}{2\pi} \, V(r) \, I_e({\bf r}) \, T({\bf r}, {\bf q_0})  \, \nonumber \\
   &  \ & - \, \frac{\mu}{2\pi} \,  V(r) \,  \int \, {\mbox{d} }{\bf r'} \,
 \, \frac{ e^{iq_0|{\bf r}-{\bf r'}|} \, T ({\bf r'}, {\bf q}_0 )\, - \, T({\bf r}, {\bf q_0}) }
 { |{\bf r}-{\bf r'} | }   \, ,
\end{eqnarray}
\noindent where
\begin{equation}
I_e({\bf r})\, =\, \int \, {\mbox{d} }{\bf r'} \,
\frac {1}{ |{\bf r}-{\bf r'} |}\, .
\end{equation}
The integral over ${\bf r'}$ in Eq. (18)  can now be approximated by a quadrature rule.
It is understood that  the term $I_e\, $ is to be calculated
analytically over the computational domain.
Introducing a cutoff $r_{max}$ for the radial variable $r$, we find that
\begin{equation}
I_e({\bf r})\, = \, \int_0^{r_{max}} r'^2\, dr' \int_{-1} ^{+1} dx' \int_0^{2\pi} d\phi' \, |{\bf r}-{\bf r'}|^{-1}
\, =\, 4\pi\,  (\frac{r_{max}^2}{2}-\frac{r^2}{6})
\end{equation}
Note that $I_e({\bf r})$ is in fact independent of the orientation of ${\bf r}$.
 The same integral
$\int_0^{r_{max}} r'^2\, dr' \int_{-1} ^{+1} dx' \int_0^{2\pi} d\phi' \, |{\bf r}-{\bf r'}|^{-1} $ also occurs
as part of the second term in the right hand side of Eq. (18), where
however it is to be evaluated via the quadrature rule
chosen for the discretization of the integral equation.

Let ${\bf r}_{\alpha}, \alpha\, =\, 1,2,...,N_Q$ be a set of quadrature points
over the computational ${\bf r}-$domain with corresponding weights denoted by $w_{\alpha}$.
We will construct this three-dimensional quadrature rule
as the direct product of single-variable quadrature rules for $r$, $x$ and $\phi$.
Let $\{r_i,\, i=1,..., N_r\},
\{ x_j, \, j=1,...,N_x \}$ and $\{ \phi_k, \, k=1,...,N_{\phi} \} $ be
the sets of quadrature points chosen for $r,x$ and $\phi$ over the intervals $[0, r_{max}]$,
$[-1,+1]$ and  $[0, 2\pi]$, with corresponding weights $\{w_{r_i} \},
\{ w_{x_j} \}$ and $\{ w_{\phi _k} \} $, respectively. Using the composite index $\alpha$ for the
index combination $(ijk) $,  the position vector whose spherical components are
$r_i$,  $x_j$ , $\phi_k$ is denoted as
${\bf r}_\alpha, \alpha=1,2,...,N_Q$ , where $N_Q=N_rN_xN_{\phi}\, $.
With this notation, the weights of the three-dimensional quadrature rule
are $w_{\alpha}= r_i^2w_{r_i}w_{x_j}w_{\phi_k}$.

Approximating the integral over  ${\bf r'}$  in Eq. (18) by the quadrature rule chosen and
collocating ${\bf r}$ at the quadrature points, we obtain
   \begin{eqnarray}
 T ({\bf r}_{\alpha}, {\bf q}_0  ) &  = &
  V(|{\bf }r_{\alpha}|) <{\bf r}_{\alpha} | {\bf q}_0 > \,  + \,
   V(|{\bf }r_{\alpha}|)\,  \,  C ({\bf r}_{\alpha}) \,
    T({\bf r}_{\alpha}, {\bf q_0}) \,   \nonumber \\
   & \ & +\,\ V(|{\bf }r_{\alpha}|)\, \,  \Sigma_{\alpha'=1}^{N_Q}\, \,
	{\bar \delta}_{\alpha \alpha'}\,
	G_{0}({\bf r}_{\alpha}, {\bf r}_{\alpha'} )\,  w_{\alpha'} \,
   T ({\bf r}_{\alpha'}, {\bf q}_0 )\, ,
\end{eqnarray}
\noindent where $\alpha=1,2,...,N_Q$ , ${\bar \delta}_{\alpha {\alpha'}}\, =
\, 1 - \delta_ {\alpha \alpha'}$, and
\begin {equation}
	{ C} ({\bf r}_{\alpha})\, =\, -\frac{\mu}{2\pi} \,  [I_e({\bf r}_{\alpha}) \, - \, I_a ({\bf r}_{\alpha})]
	\end{equation}
	\noindent Here $I_e$ is as defined in Eq. (20) and
	\begin {equation}
	I_a({\bf r}_{\alpha})\,  = \,  \Sigma_{\alpha'=1}^{N_Q}\, \,
	{\bar \delta}_{\alpha \alpha'}\, w_{\alpha'}/| {\bf r}_{\alpha} - {\bf r}_{ \alpha'}|\, .
	\end{equation}
	For a given initial momentum vector ${\bf q_0}$, Eq. (22) represents a system of $N_Q$ linear equations.
	
	Although computations with small to  moderate values of $N_r, N_x,$ and $N_\phi$
	(say, with 20-30 points in each variable) can be carried out in commonly
	 available computational platforms,  more realistic computations
	would require sophisticated programming  techniques and
	 computational environments. Therefore,  the two-variable equation with azimuthal angle
	eliminated is of  practical interest. Although the integration over $\phi$ and/or $\phi'$
	 implicit in the definition of ${\hat G}_0(r,x;r',x')$
 weakens the singularity of the Green's function $G_0({\bf r}, {\bf r'})$,
 the numerical treatment of the two-variable LS equation requires a careful handling of the kernel.
 By application to Eq. (13) of the subtraction trick used for the handling of the kernel singularity
in the full three-dimensional equation, Eq. (13) can be recast as
\begin{eqnarray}
 {\hat T}(r,x;q_0,x_0) & = & V(r)<rx|q_0x_{q_0}> \, - \,  (\mu/2\pi) \,   V(r)\, I_e(r)\, T(r,x;q_0,x_0) \,     \  \nonumber \\
         \, & + & \,  V(r) \int_{0}^{\infty} r'^2 dr' \int_{-1}^{1}  dx' \,
         \int d{\phi'}\,{\Big \{ } \, G_0({\bf r},{\bf r'})\,  T(
         r',x';q_0,x_0)    \, \nonumber \\
    \, &  \  & \ \ \ \ \ \ \ \ \ \ \ \ \ \ \ \ \ \ \ \ \
     + (\mu/2\pi) | {\bf r}-{\bf r'}| ^{-1} \, T(r,x;q_0,x_0) {\Big \} } \, .
  \end{eqnarray}
	In Eq. (24),  the azimuthal angle $\phi$ of the vector ${\bf r}$
	has been set to  zero.  This choice can be made
	because integrals
	$ \int d{\phi'}\, G_0({\bf r},{\bf r'})\, $ and  $\int d{\phi'}\, | {\bf r}-{\bf r'}| ^{-1} \, $
	are independent of the azimuthal angle of ${\bf r}$,
	as pointed out in connection with Eq. (11).
	Eq. (24)  is now in a form suitable for quadrature discretization.

	Approximating  the integrals over $r'$, $x'$ and $\phi'$ in Eq. (24) by
	the quadrature rule and collocating $r$ and $x$
	at the quadrature points  $\{ r_i, \, i=1,2,...,N_r  \}$
	and $ \{ x_j , \,  j=1,2,...,N_x \}$, respectively,
	we obtain a system of $N_rN_x$ equations:
	\begin{eqnarray}
 {\hat T}(r_i,x_j;q_0,x_0) &  = &  V(r_i)<r_ix_j|q_0x_{q_0}> \, - \,\,
 V(r_i)\,{\hat C}(r_i,x_j)\,  {\hat T}(r_i,x_j;q_0,x_{q_0})	\,    \  \nonumber \\
        \  \  & + &  V(r_i) \Sigma_{i'=1}^{N_r} \Sigma_{j' = 1}^{N_x}
   {\hat G}_0(r_i,x_j;r_{i'},x_{j'})\, r_{i'}^2  w_{r_{i'}}w_{x_{j'}}\, {\hat T}(r_{i'},x_{j'};q_0,x_{q_0}) \,   , \nonumber \\
   & \ &\,
  \end{eqnarray}	
	 where
	\begin{eqnarray}
	{\hat G}_0(r_i,x_j;r_{i'},x_{j'})\,  & = &  \, -(\mu / 2\pi) \, \Sigma_{k=1}^{N_{\phi}} \,
	w_{\phi_k}\, e^{ iq_0 d(ij,i'j';k) }\, / d(ij,i'j';k) \, , \nonumber \\
	d(ij,i'j';k) \,  & = &  \,
     [ \, r_i^2 \, + \, r_{i'}^2\,  -\, 2r_i r_{i'}\, (\, x_jx_{j'}\, +\, s_js_{j'}\,  cos\,  \phi_k\, ) \,  ]^{1/2}
		\, , \nonumber \\
    {\hat C}(r_i,x_j)\, &  = & \,  -\, (\mu/2\pi)\, \,  [\,  I_e(r_i)  \, - \,I_a(r_i,x_j)\, ] \, ,
     \nonumber  \\
    I_a(r_i,x_j) \,  &  =  & \,  \Sigma_{k=1}^{N_{\phi}} \,
	w_{\phi_k}\, / d(ij,i'j';k) \, . \nonumber
	\end{eqnarray}	
	The  singularity-correction term ${\hat C}(r_i,x_j)\, $   turns out to be  crucial for the success of the Nystrom method.\\

\noindent {\bf 4 Computational Implementation and Results }\\

\noindent To test the two- and three-variable implementations
of the Nystrom method discussed in the previous section,
 Malfliet-Tjon III ( MT-III) model for the two-nucleon potential
 has been used:
\begin{displaymath}
V(r)\, =\, V_R\, e^{-\mu_Rr}\, - \, V_A\, e^{-\mu_Ar}\,
\end{displaymath}
The parameters for MT-III potential are taken from Ref. [6]: $V_A=626.8932$ MeV fm,
$V_R=1438.723$ MeV fm, $\mu_A=1.55$ fm$^{-1}$ and $\mu_R=3.11$ fm$^{-1}$.
For the two-nucleon calculations, we set nucleon mass
and $\hbar$ to unity and take $fm$ as the unit of length.
The nucleon mass adopted yields the conversion factor
$1 fm^{-2}=41.47$ MeV.\\

The cut-off  $r_{\max}$ for the $r$-variable is taken as $ 15\, fm $, although a value of $8 \, fm$ is sufficient
for about  4 digit accuracy.  Quadrature grid for $r$ is uneven: denser for small $r$, coarser for large $r$.
The interval $[0,r_{max}]$ is divided  into 4 sub-intervals: $[0,0.5], [0.5, 2], [2,10], [10,15] $,
which are in turn
 subdivided into $I_i$ -elements, $i=1,2,3,4$.  Each  element is mapped to $[-1,+1]$, and
a set of $n_r$ Gauss-Legendre points
and their corresponding weights are generated for each element. By combining
the quadrature  points  and weights
for all elements,  a composite quadrature rule of  $N_r$ points is generated.
Here $N_r\, =\, I_r n_r$, with $I_r \, (\equiv I_1+I_2+I_3+I_4)$
 denoting the total number of $r$-elements . Similarly,
the interval $ [-1,+1]$ for the $x$-variable was divided into $I_x$ equal elements, with $n_x$
Gauss-Legendre points chosen in each element. Thus, a composite quadrature rule with
$N_x\, =\,  I_x n_x$ was generated.  For doing the $\phi$ integrals,
the  interval $[0,2 \pi]$  was divided into $I_{\phi}$ equal elements,
with $n_{\phi}$ Gauss-Legendre points in each element, yielding a composite quadrature rule with
$N_{\phi}= I_{\phi}n_{\phi}$. \\

Reference results for the MT-III potential
were obtained with  momentum-space Nystrom calculations,
as reported in Ref. [13],
and are stable within seven digits after the decimal point to further
variations in computational parameters .\\

\noindent {\bf 4.1 Results of three-dimensional calculations }\\

\noindent Tables 1 and 2 report results of Nystrom calculations
with different values of $N_r$, $N_x$ and
$N_{\phi}$.  Table 1 probes the convergence
with respect to $N_r$ with fixed $N_x$ and $N_{\phi}$,
while Table 2 with respect to $N_x$ and $N_{\phi}$ with fixed $N_r$ .

The largest calculation in Table 1  is for $N_r=100$ and corresponds to
the following distribution of quadrature points for $r$: $I_1=6, I_2=6, I_3=9, I_4=4,$
and $n_r=4$.  With $N_x=20$ and $N_{\phi}=20$,
this corresponds to $N_Q=40 000$.
As the matrices of this  size  can not be kept
in the fast memory, they  are stored
in the disk space in a block-by-block fashion.
 The  system of equations is  solved  either by
   a direct out-of-core equation solver  (a block-by-block scheme of
	Gaussian elimination with partial pivoting
	that was described earlier in [21])
   or by  Pade re-summation of the Born series
   generated from Eq. (21). Typically, $[8,7]$-approximant is sufficient  for convergence.
   It is reassuring that direct and Pade solutions agree within  at least 8 significant digits 
   for a given set of
   computational parameters.

   As we were restricted to use  rather modest values
	($12$ to $20$)  for $N_x$ and $N_{\phi}$ to avoid excessively large $N_Q$,
   accuracy of the results of three dimensional calculations is
    limited to 3-4 significant figures.
    For the  largest three-dimensional calculation  in Table 1
		(namely, the run with $N_r=100, N_x=N_{\phi}=20$ and hence $N_Q=40000$),
	 the largest relative error among  the transition matrix elements
   reported in Table 1 is about 0.3 percent.
   Although the value of $N_r=80-100$ may be close to being  adequate for the $r$-variable,
   Table 2 shows that the values of $N_x$ and $N_{\phi}$ are far from being sufficient.
   Indeed, two-dimensional calculations of next section suggest that $N_x$ and $N_{\phi}$ must be
   in the order of $60$ to $80$ if we ask for results accurate
	to within 5-6 significant figures.
   However, going beyond $N_Q=40000$ is a formidable task,
   requiring special programming  and   hardware, and
	has not been attempted. Instead,
   two-dimensional reduced equations were solved
   to obtain results with  7-8 significant figures.\\

\noindent {\bf 4.2 Results of two-variable calculations }\\

\noindent Tables 3 and 4 report convergence of the Nystrom method
for the two-variable LS equation with
respect to $N_r$ and $N_x$. Integration over $\phi$
implicit in the calculation of
${\hat G}(r,x;r',x')$ is done with $N_{\phi}=64$, which is sufficient for
the stability of results  to the number of digits shown  in these tables.
For the most refined calculation
(with $N_r=440$ and $N_x=96$ ) in Table 3,
the largest absolute deviation (from the reference solution )
 is $1.4 \times 10^{-6}$ for the imaginary
part of the forward amplitude at 400 MeV.
This corresponds to a relative error of about
$2.3 \times 10^{-5}$ percent.
 For the other values of $x$ in the same calculation,
 the absolute value of the  difference
between momentum-space and coordinate-space calculations
is less than $5 \times 10^{-7}$, while  relative errors
 are less than $ 5 \times 10^{-4}$ percent.

  The relatively high values of $N_r$ and $N_x$
were needed  to achieve agreement within 7-8 significant figure.
However, more moderate values like $N_r=176$ and $N_x=60$ would
be sufficient to obtain
agreement within  5-6 significant figures.
We note in passing that, reference results obtained
from momentum-space Nystrom method [11,13]
involves about the same level of computational effort
as the coordinate-space Nystrom method
for comparable levels of convergence .\\

Finally we would like to point out the role that
the singularity correction  term ${\hat C}(r_i,x_j)\, $plays
in the   performance of the coordinate-space
 Nystrom method for the reduced LS equation.
Table 5 gives the results obtained  by setting
${\hat C}(r_i,x_j)\, =0$ in Eq. (25) ,
which corresponds to pretending as if the kernel has no singularity.
As the singularity of the reduced kernel is
in fact weaker, ignoring it does not lead
to a catastrophy, but results are of low accuracy.
Note, however,  that the correction term
 ${ C} ({\bf r}_{\alpha})\, $ in (21)
for the three-dimensional case
 plays a much more crucial role,
 for without it Nystrom idea is totally inapplicable.\\

\noindent {\bf 5 Conclusions }\\

As part of our continuing  interest  in multi-variable methods
for solving scattering integral equations
without invoking  expansions over angular-momentum states,
 we have considered the LS integral equation in coordinate space.
 Both the original three-dimensional and
 the reduced two-variable versions have been considered.
 To apply the Nystrom method
 (which combines  quadrature discretization with  collocation),
 a suitable scheme for handling
 the (moving -type) singularity of the free Green's function
 has  been implemented.
 The basic idea is to cancel out the Green's function singularity
 by subtracting  a known singularity (namely $|{\bf r}={\bf r'}|^{-1}$) which  can be integrated
 in closed form, leaving a smoother kernel that can be integrated by quadrature.
 Of course, the kernel of the momentum-space LS equation is also singular,
 but the singularity occurs
 at a fixed value of the integration variable
  and is somewhat easier
  to treat by a similar subtraction  scheme
  as discussed, e.g., in [11]. Apart from this small difference
  in handling the kernel singularities, momentum-space and coordinate-space
   Nystrom methods involve about the same level of computational effort.
   For local potentials, however, coordinate-space approach may be more natural,
    as the need for the (possibly numerical) calculation 
    of the momentum-space representation is avoided. 
    For instance, 
    in the context of  a Faddeev-equations approach to
    three-atom problems  [24-25],  
    calculation of  the atom-atom transition matrices 
    (for  numerically available diatomic potentials) 
    would be more practical in coordinate space.

Calculations presented in this article have been done with complex arithmetic.
The  K-matrix version of the
 present approach is possible and could lead to some computational savings  (in
computation time and memory needs) by
allowing to work in real arithmetic. However, obtaining T-matrix from K-matrix would involve
the solution of an additional integral equation in the angular variables.  This possibility is currently under
 investigation.

The present calculations show that the Nystrom method coupled
with  the particular singularity removal scheme adopted
 is a viable procedure
that is capable of producing accurate solutions
of the LS equation. However, matrix dimensions in the Nystrom method implemented
with direct-product quadrature schemes quickly becomes computationally
prohibitive. In fact, both coordinate- and momentum-space versions 
suffer from this problem.  Variational  methods based on multivariate bases
may provide the alternative to the multivariate Nystrom method. 
The singularity subtraction
scheme used in the present article would also
be applicable in calculating the matrix elements that come up
in the variational approaches and in other Galerkin-Petrov methods.
 However the so-called  "curse of dimensionality"
hampers all methods that make use of direct-product  bases. 
The radial basis function (rbf)
approach  (which is nearly dimension independent)
 has emerged  during recent years as
an alternative to direct-product bases [22,23].
In a recent article [13] we explored the use of rbf's (in momentum space))
in relation to momentum-space LS equation
with promising results (for both three-dimensional and two-dimensional versions).
A logical continuation of the present work 
would be to consider rbf expansions (in coordinate space)
as a means of solving
the coordinate-space LS equation and of obtaining separable expansions of the
multivariable T-matrix. Separable expansions (of manageable rank) 
in multivariate bases 
for the two-particle T-matrix would be particularly useful for
three-body calculations  without angular momentum
decomposition [2,14,15].   \\

\noindent {\bf References }

\begin{enumerate}
\item Kadyrov, A.S., Abdurakhmanov, I.B., Bray, I.,
Stelbovics, A.T.:
Three-dimensional integral-equation approach to proton- and
antiproton-hydrogen collisions.
Phys. Rev. A {\bf 80}, 022704 (2009)

\item Elster, Ch., Gl\"ockle, W., Wita{\l}a, H.:
A new approach to the 3D Faddeev equation for three-body scattering.
Few-Body Syst. {\bf 45}, 1 (2009)

\item Elster, Ch., Thomas, J.H., Gl\"ockle, W.:
Two-body T-matrices without angular-momentum decomposition: Energy
and momentum dependences. Few-Body Syst. {\bf 24}, 55 (1998)

\item Shertzer, J., Temkin, A.: Direct calculation of
the scattering amplitude without-partial wave analysis.
Phys. Rev. A {\bf 63}, 062714 (2001)

\item Caia, G.L., Pascalutsa, V., Wright, L.E.: Solving potential
scattering equations without partial wave decomposition.
Phys. Rev. C {\bf 69}, 034003 (2004)

\item Kessler, B.M., Payne, G.L., Polyzou, W.N.: Application of
wavelets to singular integral scattering equations.
Phys. Rev. C {\bf 70}, 034003 (2004)

\item Kadyrov, A.S., Bray, I., Stelbovics, A.T., Saha, B.: Direct
solution
of the three-dimensional Lippmann-Schwinger equation.
J. Phys. B {\bf 38}, 509 (2005)

\item Ramalho, G., Arriaga, A., Pe\~na, M.T.: Solution of the
spectator
equation for relativistic NN scattering without partial
wave expansion. Few-Body Syst. {\bf 39}, 123 (2006)

\item Rodr{\'i}guez-Gallardo, M., Deltuva, A., Cravo, E., Crespo,
R.,
Fonseca, A.C.: Two-body scattering without angular-momentum
decomposition. Phys. Rev. C {\bf 78}, 034602 (2008)

\item Veerasamy, S., Elster, Ch., Polyzou, W.N.:
Two-nucleon scattering without partial waves using
a momentum space Argonne V18 interaction.
Few-Body Syst. {\bf 54}, 2207 (2012)

\item Kuruo\u{g}lu, Z.C.: Weighted-residual methods for the solution of
two-particle Lippmann-Schwinger equation without partial-wave decomposition.
Few-Body Syst. {\bf 55}, 69 (2014)

\item Kuruo\u{g}lu, Z.C.: Bateman method for
two-body scattering without partial-wave decomposition.
J. Math. Chem. {\bf 52},  1857  (2014)

\item Kuruo\u{g}lu, Z.C.: Finite-rank multivariate-basis expansions of the resolventoperator as a means of solving multivariable Lippmann-Schwinger
equation for two-particle scattering.
Few-Body Syst. {\bf  55}, 1167 (2014)

\item Schadow, W., Elster, Ch., Gl\"ockle, W.:
Three-body scattering below breakup threshold: An approach without
using partial waves. Few-Body Syst. {\bf 28}, 15 (2000)

\item Liu, H., Elster, Ch., Gl\"ockle, W.:
Three-body scattering at intermediate energies.
Phys. Rev. C {\bf 72}, 054003 (2005)

\item Staszewska, G., Truhlar D.G.: Convergence of $L^2$ methods for scattering problems.
J. Chem. Phys. {\bf 86}, 2793 (1987)

\item Sun,Y., Kouri, D.J., Truhlar, D.G.: A comparative analysis of variational methods
for inelastic and reactive scattering. Nucl. Phys. A {\bf 508}, 41c (1990)

\item Atkinson, K.E.: A Survey of Numerical Methods for the Solution of
Fredholm Integral Equations of the Second Kind. SIAM, Philadelphia (1976)

\item Sheng,X-Q, Song, W.: Essentials of Computational Electromagnetics.
Wiley, Singapore (2012)

\item Sadiku, M. N.O.,: Numerical Techniques in Electromagnetics., 2nd edn.
CRC Press, Boca Raton (2000)

 \item Kuruoglu, Z.C.,   Micha, D.A.:
  Collision dynamics of three interacting atoms: Model
  calculations of H + H$_2$ resonances.
 J. Chem. Phys. {\bf 80}, 4262 (1984).

 \item Cheney, E.W.: Multivariate Approximation Theory: Selected Topics.
  SIAM, Philadelphia (1986)

  \item Cheney, W., Light, W.: A Course in Approximation Theory. AMS, Providence (2009)

\item Kuruo\u{g}lu, Z.C.,   Micha, D.A.:
  Collision dynamics of three interacting atoms:
 The Faddeev equations in a diabatic electronic basis.
 J. Chem. Phys. {\bf 79}, 6115 (1983)
 
 \item    Micha, D.A., Kuruo\u{g}lu, Z.C.,  : Atom-diatom resonances within a many-body approach to
reactive scattering.  
 ACS Symposium Series, {\bf 263}, 401 (1984).
 
 \item Kuruo\u{g}lu, Z.C.,   Micha, D.A.:
   Calculation of resonances in the H + H$_2$ reaction
  using the Faddeev-AGS method. Int. J. Quantum. Chem.  {\bf S23}, 105 ( 1989)

\end {enumerate}

\newpage

\noindent {\bf Table 1}  Convergence study for the three-dimensional Nystrom method
with respect to the number of quadrature points $N_r$ for the  $r$-variable.
 Listed  are the on-shell T-matrix elements
 $<q_0x\phi|T(E)|q_0x_0\phi_0>\, $ with $x_0=1.0$ and
 $\phi=\phi_0=0$ at $E=150$ and $E=400$ MeV obtained with different values of $N_r$.
 The number of quadrature points used  for the angular variables
are $N_x=20$ and $N_{\phi}=20$.\\

\begin{tabular}{  l l c l l l l}
\hline
 \multicolumn{1}{l}{}
 & \multicolumn{3}{l} {Re $<q_0x\phi| T|q_0x_0\phi_0>  $}
 & \multicolumn{3}{l} { Im $<q_0x\phi|T|q_0x_0\phi_0>$}\\
 \hline
\multicolumn{1}{l}{$N_r$}
  &\multicolumn{1}{l}{x=1.0 }
  & \multicolumn{1}{c} {x=0.0 }
  & \multicolumn{1}{l}{x=-1.0  }
  &\multicolumn{1}{l}{ x=1.0 }
  & \multicolumn{1}{l} {x=0.0 }
	& \multicolumn{1}{l}{ x=-1.0 }\\
\hline \hline
\multicolumn{2}{l}{}
& \multicolumn{1}{c}{$E=150$ MeV}
& \multicolumn{4}{l}{ } \\
\hline
 64  &  -6.0976 &  0.4923 & 0.2351  &   -1.9294  & 0.2864 & 0.3656\\
 72   & -6.0963 &  0.4923 & 0.2349  &   -1.9305  & 0.2864 & 0.3656\\
  80   & -6.0955 &  0.4922& 0.2347   &    -1.9312  & 0.2864 & 0.3655\\
  88  & -6.0954 &  0.4922& 0.2347   &    -1.9315  & 0.2864 & 0.3656\\
 100  & -6.0949 &  0.4922 & 0.2346   &   -1.9319  & 0.2864 & 0.3655\\	
 \hline
 \multicolumn{1}{l}{Ref. [11] }
& \multicolumn{1}{l}{-6.0928}
& \multicolumn{1}{c}{0.4918}
&\multicolumn{1}{l}{0.2340 }
& \multicolumn{1}{l}{-1.9373}
& \multicolumn{1}{l}{0.2861}
& \multicolumn{1}{l}{0.3657}\\
	\hline \hline
\multicolumn{2}{c}{}
& \multicolumn{1}{c}{$E=400$ MeV}
&\multicolumn{4}{l}{ } \\
\hline
 64 &  -6.1699  & 0.4556 & 0.2498  &   -1.2982  & 0.1110 & -0.0785\\
 72   &  -6.1685  & 0.4554 & 0.2496  &   -1.3005  & 0.1108 & -0.0784\\
  80   & -6.1677   & 0.4553 & 0.2495  &   -1.3018  & 0.1108 & -0.0783\\
  88  & -6.1675  & 0.4553 & 0.2495  &   -1.3022  &  0.1108 & -0.0783\\
 100    & -6.1670  & 0.4552 & 0.2495  &    -1.3031 &  0.1108 & -0.0782\\
\hline
 \multicolumn{1}{l}{Ref. [11]}
  &\multicolumn{1}{l}{-6.1638}
	& \multicolumn{1}{c}{0.4549}
& \multicolumn{1}{l}{0.2491}
& \multicolumn{1}{l}{-1.3116}
 & \multicolumn{1}{l}{0.1108}
& \multicolumn{1}{l}{-0.0776}\\
\hline \hline
\end{tabular}

\newpage
\noindent {\bf Table 2}  Convergence study for the three-dimensional Nystrom method
with respect to the number of quadrature points $N_x$ and $N_{\phi}$
 in the $x $ and $\phi$ variables, respectively.
 Listed  are the on-shell T-matrix elements
 $<q_0x\phi|T(E)|q_0x_0\phi_0>\, $ with $x_0=1.0$ and $\phi=\phi_0=0$
 at $E=150$ and $E=400$ MeV.
  For calculations of this table,  $N_r=100$. \\

\begin{tabular}{  l l l c l l l l}
\hline
 \multicolumn{2}{l}{}
 & \multicolumn{3}{c} {Re $<q_0x\phi| T|q_0x_0\phi_0>  $}
 & \multicolumn{3}{c} { Im $<q_0x\phi|T|q_0x_0\phi_0>$}\\
 \hline
 \multicolumn{1}{l}{$N_x$}
& \multicolumn{1}{l}{$N_{\phi} $}
  &\multicolumn{1}{l}{x=1.0 }
  & \multicolumn{1}{c} {x=0.0 }
  & \multicolumn{1}{l}{x=-1.0  }
  &\multicolumn{1}{l}{ x=1.0 }
  & \multicolumn{1}{l} {x=0.0 }
	& \multicolumn{1}{l}{ x=-1.0 }\\
\hline \hline
\multicolumn{3}{l}{}
& \multicolumn{1}{c}{$E=150$ MeV}
& \multicolumn{4}{l}{ } \\
\hline

 12 & 12 & -6.0939 &  0.4982 & 0.2357  &    -1.9198  & 0.2870 & 0.3648\\	
  16 & 16 & -6.0951 &  0.4931 & 0.2352   &   -1.9284  & 0.2865 & 0.3654\\	
  20 & 20 & -6.0949 &  0.4922 & 0.2346   &   -1.9319  & 0.2864 & 0.3655\\	
 \hline
 \multicolumn{2}{l}{Ref. [11] }
& \multicolumn{1}{l}{-6.0928}
& \multicolumn{1}{c}{0.4918}
&\multicolumn{1}{l}{0.2340 }
& \multicolumn{1}{l}{-1.9373}
& \multicolumn{1}{l}{0.2861}
& \multicolumn{1}{l}{0.3657}\\
	\hline \hline
\multicolumn{3}{c}{}
& \multicolumn{1}{c}{$E=400$ MeV}
&\multicolumn{4}{l}{ } \\
\hline
 12 & 12  & -6.1722  & 0.4565 & 0.2593  &    -1.2893  & 0.1089 & -0.0793\\
  16 & 16  & -6.1685  & 0.4579 & 0.2490  &    -1.2985 &  0.1108 & -0.0785\\
  20 & 20  & -6.1670  & 0.4552 & 0.2495  &    -1.3031 &  0.1108 & -0.0782\\
\hline
 \multicolumn{2}{l}{Ref. [11]}
  &\multicolumn{1}{l}{-6.1638}
	& \multicolumn{1}{c}{0.4549}
& \multicolumn{1}{l}{0.2491}
& \multicolumn{1}{l}{-1.3116}
 & \multicolumn{1}{l}{0.1108}
& \multicolumn{1}{l}{-0.0776}\\
\hline \hline
\end{tabular}

\newpage
\noindent {\bf Table 3}  Convergence study for the two-variable Nystrom method
with respect to the number of quadrature points $N_r$ for the  $r$-variable.
 Listed  are the on-shell T-matrix elements $<q_0x|{\hat T}(E)|q_0x_0>\, $ with $x_0=1.0$
 at $E=150$ and $E=400$ MeV for different values of $N_r$.  For calculations reported in this table,
the number $N_x$ of quadrature points for the $x$-variable is $96$ .\\

\begin{tabular}{  l l l l l l l}
\hline
 \multicolumn{1}{l}{}
 & \multicolumn{3}{c} {Re $<q_0x| {\hat T|}q_0x_0>  $}
 & \multicolumn{3}{c} { Im $<q_0x|{\hat T}|q_0x_0>$}\\
 \hline
 \multicolumn{1}{l}{$N_r$}
  &\multicolumn{1}{l}{x=+1.0 }
  & \multicolumn{1}{l} {x=0.0 }
  & \multicolumn{1}{l}{x=-1.0  }
  &\multicolumn{1}{l}{ x=+1.0 }
  & \multicolumn{1}{l} {x=0.0 }
	& \multicolumn{1}{l}{ x=-1.0 }\\
\hline \hline
\multicolumn{3}{l}{} & \multicolumn{1}{l}{$E=150$ MeV} &
\multicolumn{3}{l}{ } \\
\hline
 44     & -6.0934081  &  0.4929036 & 0.2316417  & -1.935842 & 0.2869646 & 0.3608944\\
 88     & -6.0928178 & 0.4917631 & 0.2339522  &   -1.9372708 & 0.2860977 & 0.3656473\\
  132   & -6.0927905 & 0.4917664& 0.2339563   &   -1.9372523  & 0.2860970 & 0.3656480\\
  176   & -6.0927853 & 0.4917668 & 0.2339572   &   -1.9372487 & 0.2860967 & 0.3656484\\
 220   &  -6.0927834 & 0.4917670 & 0.2339575   &   -1.9372474 & 0.2860967 & 0.3656485\\	
 264   & -6.0927829   & 0.4917669 & 0.2339575   &  -1.9372471 & 0.2860966 & 0.3656485\\	
 308   & -6.0927826   & 0.4917669 & 0.2339575   &   -1.9372469 & 0.2860966 & 0.3656485\\	
 352   & -6.0927825   & 0.4917669  & 0.2339575   &   -1.9372468 & 0.2860965 & 0.3656485\\	
 440   & -6.0927823   & 0.4917669 & 0.2339575   &   -1.9372467 & 0.2860965 & 0.3656485\\	
 \hline
 \multicolumn{1}{l}{Ref. [11] }  &
\multicolumn{1}{l}{-6.0927820} & \multicolumn{1}{c}{0.4917677} &
\multicolumn{1}{l}{0.2339576 }  &
\multicolumn{1}{l}{-1.9372472} &
\multicolumn{1}{l}{0.2860968} & \multicolumn{1}{l}{0.3656489}\\
	\hline \hline
\multicolumn{3}{l}{} & \multicolumn{1}{l}{$E=400$ MeV} &
\multicolumn{3}{l}{ } \\
\hline
  44     & -6.1645448    & 0.4540738  &  0.2350731   &   -1.3147435   &  0.1087639   &   -0.0891877\\
  88     & -6.1638597   & 0.4549345   &  0.2490815   &   -1.3116647   &  0.1107643    &  -0.0776850\\
  132   &-6.1638233    & 0.4549278   &  0.2491374   &   -1.3116464   &  0.1107539    &  -0.0776415\\
  176   &-6.1638145    & 0.4549285   &  0.2491381   &   -1.3116431   &  0.1107532    &   -0.0776417\\
 220    &-6.1638115    & 0.4549287   &  0.2491384   &   -1.3116420   &  0.1107529    &   -0.0776418\\
264     &-6.1638105    & 0.4549287   &  0.2491384   &   -1.3116416   &  0.1107528    &   -0.0776418\\
308     &-6.1638100    & 0.4549287   &  0.2491384   &   -1.3116414   &  0.1107528    &   -0.0776417\\
352     &-6.1638097    & 0.4549287  &  0.2491384    &   -1.3116413   &  0.1107528    &   -0.0776417\\
440     &-6.1638094    & 0.4549288  &  0.2491385    &   -1.3116411   &  0.1107528    &   -0.0776417\\

\hline
 \multicolumn{1}{l}{ref. [11]}
  &\multicolumn{1}{l}{-6.1638080} & \multicolumn{1}{c}{0.4549298}
& \multicolumn{1}{l}{0.2491389 }
& \multicolumn{1}{l}{-1.3116411}
 & \multicolumn{1}{l}{0.1107532} & \multicolumn{1}{l}{-0.0776420}\\
\hline \hline
\end{tabular}

\newpage
\noindent {\bf Table 4}  Convergence of the Nystrom solution of the  two-variable LS equation
with respect to $N_x$, the number of quadrature points in $x$-variable.
 Listed  are the on-shell T-matrix elements $<q_0x|{\hat T}(E)|q_0x_0>\, $ with $x_0=1.0$
 at $E=150$ and $E=400$ MeV for different values of $N_x$.  For calculations reported in this table,
the number $N_r$ of quadrature points for the $r$-variable is $352$ .\\

\begin{tabular}{  l l l l l l l}
\hline
 \multicolumn{1}{l}{}
 & \multicolumn{3}{c} {Re $<q_0x| {\hat T|}q_0x_0>  $}
 & \multicolumn{3}{c} { Im $<q_0x|{\hat T}|q_0x_0>$}\\
 \hline
 \multicolumn{1}{l}{$N_x$}
  &\multicolumn{1}{l}{x=+1.0 }
  & \multicolumn{1}{l} {x=0.0 }
  & \multicolumn{1}{l}{x=-1.0  }
  &\multicolumn{1}{l}{ x=+1.0 }
  & \multicolumn{1}{l} {x=0.0 }
	& \multicolumn{1}{l}{ x=-1.0 }\\
\hline \hline
\multicolumn{3}{l}{} & \multicolumn{1}{l}{$E=150$ MeV} &
\multicolumn{3}{l}{ } \\
\hline
 40     & -6.0927853  &  0.4917625 & 0.2339612  &  -1.9372358 & 0.2860970 & 0.3656424\\
 60    & -6.0927823   & 0.4917658 & 0.2339585   &   -1.9372437 & 0.2860958 & 0.3656472\\
  80   & -6.0927823   & 0.4917666 & 0.2339578   &   -1.9372460  & 0.2860965 & 0.3656482\\
	 96   & -6.0927825   & 0.4917669 & 0.2339575   &   -1.9372468  & 0.2860965 & 0.3656485\\
  108  &  -6.0927825 & 0.4917670 & 0.2339574    &   -1.9372471 & 0.2860966 & 0.3656486\\
  120   &  -6.0927826 & 0.4917671 & 0.2339574   &   -1.9372473 & 0.2860966 & 0.3656487\\	
 \hline
 \multicolumn{1}{l}{Ref. [11] }  &
\multicolumn{1}{l}{-6.0927820} & \multicolumn{1}{c}{0.4917677} &
\multicolumn{1}{l}{0.2339576 }  &
\multicolumn{1}{l}{-1.9372472} &
\multicolumn{1}{l}{0.2860968} & \multicolumn{1}{l}{0.3656489}\\
	\hline \hline
\multicolumn{3}{l}{} & \multicolumn{1}{l}{$E=400$ MeV} &
\multicolumn{3}{l}{ } \\
\hline
  40     & -6.1638379    & 0.4549246  &  0.2491048   &   -1.3116480  &  0.1107446   &   -0.0776408\\
  60     & -6.1638144    & 0.4549259  &  0.2491388   &   -1.3116420  &  0.1107508   &   -0.0776415\\
  80     &-6.1638106    & 0.4549281   &  0.2491383  &   -1.3116414   &  0.1107523    &  -0.0776417\\
  96     &-6.1638097    & 0.4549287   &  0.2491384   &   -1.3116413   &  0.1107528    &   -0.0776417\\
 108   &-6.1638095    & 0.4549290   &  0.2491385   &   -1.3116413   &  0.1107530     &   -0.0776417\\
 120     &-6.1638093    & 0.4549291   &  0.2491385   &   -1.3116413   &  0.1107531    &   -0.0776418\\

\hline
 \multicolumn{1}{l}{ref. [11]}
  &\multicolumn{1}{l}{-6.1638080} & \multicolumn{1}{c}{0.4549298}
& \multicolumn{1}{l}{0.2491389 }
& \multicolumn{1}{l}{-1.3116411}
 & \multicolumn{1}{l}{0.1107532} & \multicolumn{1}{l}{-0.0776420}\\
\hline \hline
\end{tabular}

\newpage
\noindent {\bf Table 5}  Comparison of two-variable Nystrom calculations
with and without the singularity correction term.
 Listed  are the on-shell T-matrix elements $<q_0x|{\hat T}(E)|q_0x_0>\, $ with $x_0=1.0$
 at $E=150$ and $E=400$ MeV .  For calculations in  this table, $N_r=176, \, N_x=80$.\\

\begin{tabular}{  l l l l l l l}
\hline
 \multicolumn{1}{l}{}
 & \multicolumn{3}{c} {Re $<q_0x| {\hat T|}q_0x_0>  $}
 & \multicolumn{3}{c} { Im $<q_0x|{\hat T}|q_0x_0>$}\\
 \hline
 \multicolumn{1}{l}{$\ $}
  &\multicolumn{1}{l}{x=+1.0 }
  & \multicolumn{1}{l} {x=0.0 }
  & \multicolumn{1}{l}{x=-1.0  }
  &\multicolumn{1}{l}{ x=+1.0 }
  & \multicolumn{1}{l} {x=0.0 }
	& \multicolumn{1}{l}{ x=-1.0 }\\
\hline \hline
\multicolumn{3}{l}{} & \multicolumn{1}{l}{$E=150$ MeV} &
\multicolumn{3}{l}{ } \\
\hline
 without ${\hat C}$     & -6.092514  &  0.492374 & 0.233732  &  -1.940060 & 0.286935 & 0.365925\\
 with  ${\hat C}$  & -6.092785 & 0.491767 & 0.233957   &   -1.937248 & 0.286097 & 0.365648\\
 \hline
 \multicolumn{1}{l}{Ref. [11] }  &
\multicolumn{1}{l}{-6.092782} & \multicolumn{1}{l}{0.491768} &
\multicolumn{1}{l}{0.233958 }  &
\multicolumn{1}{l}{-1.937247} &
\multicolumn{1}{l}{0.286097} & \multicolumn{1}{l}{0.365649}\\
	\hline \hline
\multicolumn{3}{l}{} & \multicolumn{1}{l}{$E=400$ MeV} &
\multicolumn{3}{l}{ } \\
\hline
 without ${\hat C}$    & -6.166331   & 0.455116 &  0.249336  &   -1.313334  &  0.111257   &   -0.077629\\
  with ${\hat C} $   & -6.163816    & 0.454928 &  0.249138  &   -1.311644  &  0.110753   &   -0.077642\\
\hline
 \multicolumn{1}{l}{ref. [11]}
  &\multicolumn{1}{l}{-6.163808} & \multicolumn{1}{l}{0.454930}
& \multicolumn{1}{l}{0.249139 }
& \multicolumn{1}{l}{-1.311641}
 & \multicolumn{1}{l}{0.110753} & \multicolumn{1}{l}{-0.077642}\\
\hline \hline
\end{tabular}

\end {document}